\documentclass[apj]{emulateapj}
\usepackage{mathptmx}  
\shorttitle{Study of the UTA feature in AGN}
\shortauthors{H. Netzer}

\newcommand  \kms      {\ifmmode {\rm km\,s}^{-1} \else km\,s$^{-1}$\fi}

\newcommand  \cmii     {\hbox{cm$^{-2}$}}
\newcommand  \ergs     {\ifmmode {\rm ergs\,s}^{-1} \else ergs s$^{-1}$\fi}
\newcommand  \ergcms   {\ifmmode {\rm ergs\,cm}^{-2}\,{\rm s}^{-1}
                        \else ergs\,cm$^{-2}$\,s$^{-1}$\fi}
\newcommand  \ergcmsA  {\ifmmode{\rm ergs\,cm}^{-2}\,{\rm s}^{-1}\,{\rm\AA}^{-1}
                        \else ergs\,cm$^{-2}$\,s$^{-1}$\,\AA$^{-1}$\fi}
\newcommand  \ergcmsHz {\ifmmode{\rm ergs\,cm}^{-2}\,{\rm s}^{-1}\,{\rm Hz}^{-1}
                        \else ergs\,cm$^{-2}$\,s$^{-1}$\,Hz$^{-1}$\fi}
\newcommand  \phcms    {\ifmmode {\rm ph\,cm}^{-2}\,{\rm s}^{-1}
                        \else ,ph\,cm$^{-2}$\,s$^{-1}$\fi}
\newcommand  \phcmsA   {\ifmmode {\rm ph\,cm}^{-2}\,{\rm s}^{-1}\,{\rm\AA}^{-1}
                        \else ph\,cm$^{-2}$\,s$^{-1}$\,\AA$^{-1}$\fi}      
\def\micron{\ifmmode \mu{\rm m} \else $\mu$m\fi}
\def\kms{\ifmmode {\rm km\,s}^{-1} \else km\,s$^{-1}$\fi}
\def\Hubble{\ifmmode {\rm km\,s}^{-1}\,{\rm Mpc}^{-1}
        \else km\,s$^{-1}$\,Mpc$^{-1}$\fi}
\def\ergsec{\ifmmode {\rm ergs\;s}^{-1} \else ergs s$^{-1}$\fi}
\def\ergscm{\ifmmode {\rm ergs\,s}^{-1}\,{\rm cm}^{-2}
          \else ergs\,s$^{-1}$\,cm$^{-2}$\fi}
\def\ergscmA{\ifmmode {\rm ergs\,s}^{-1}\,{\rm cm}^{-2}\,{\rm \AA}^{-1}
          \else ergs\,s$^{-1}$\,cm$^{-2}$\,\AA$^{-1}$\fi}
\def\ergscmHz{\ifmmode {\rm ergs\,s}^{-1}\,{\rm cm}^{-2}\,{\rm Hz}^{-1}
          \else ergs\,s$^{-1}$\,cm$^{-2}$\,Hz$^{-1}$\fi}
\def\Msun{\ifmmode M_{\odot} \else $M_{\odot}$\fi}
\def\Lsun{\ifmmode L_{\odot} \else $L_{\odot}$\fi}
\def\qo{\ifmmode q_{0} \else $q_{0}$\fi}
\def\Ho{\ifmmode H_{0} \else $H_{0}$\fi}
\def\ho{\ifmmode h_{0} \else $h_{0}$\fi}
\def\qo{\ifmmode q_{0} \else $q_{0}$\fi}
\def\ao{\ifmmode a_{0} \else $a_{0}$\fi}
\def\to{\ifmmode t_{0} \else $t_{0}$\fi}
\def\Halpha{\ifmmode {\rm H}\alpha \else H$\alpha$\fi}
\def\Hbeta{\ifmmode {\rm H}\beta \else H$\beta$\fi}
\def\hb{\ifmmode {\rm H}\beta \else H$\beta$\fi}
\def\Hgamma{\ifmmode {\rm H}\gamma \else H$\gamma$\fi}
\def\Hdelta{\ifmmode {\rm H}\delta \else H$\delta$\fi}
\def\Lya{\ifmmode {\rm Ly}\alpha \else Ly$\alpha$\fi}
\def\Lyb{\ifmmode {\rm Ly}\beta \else Ly$\beta$\fi}
\def\hi{\ifmmode \mbox{{\rm H}\,{\sc i}} \else H\,{\sc i}\fi}

\def\ciii{\ifmmode {\rm C}\,{\sc iii} \else C\,{\sc iii}\fi}

\def\o5007{[O\,{\sc iii}]\,$\lambda5007$}
\def  \kms         {\hbox{km s$^{-1}$}}          
\def  \ergs        {\hbox{erg s$^{-1}$}}              

\def  \cmii        {\hbox{cm$^{-2}$}}

\def  \etal        {{\rm et al.}}
\def  \La          {\ifmmode {\rm Ly}\alpha \else Ly$\alpha$\fi}
\def  \Ka          {\ifmmode {\rm K}\alpha \else K$\alpha$\fi}
\def  \Lb          {\ifmmode {\rm L}\beta \else L$\beta$\fi}
\def  \Ha          {\ifmmode {\rm H}\alpha \else H$\alpha$\fi}
\def  \Hb          {\ifmmode {\rm H}\beta \else H$\beta$\fi}
\def  \Pa          {\ifmmode {\rm P}\alpha \else P$\alpha$\fi}
\def  \CIIIb       {\ifmmode {\rm C}\,{\sc iii]}\,\lambda1909
                     \else C\,{\sc iii]}\,$\lambda1909$\fi}
\def  \CIV         {\ifmmode {\rm C}\,{\sc iv}\,\lambda1549
                     \else C\,{\sc iv}\,$\lambda1549$\fi}
\def  \MgII         {\ifmmode {\rm Mg}\,{\sc ii}\,\lambda2798
                     \else Mg\,{\sc ii}\,$\lambda2798$\fi}
\def  \OVI         {\ifmmode {\rm O}\,{\sc vi}\,\lambda1035
x
                     \else O\,{\sc vi}\,$\lambda1035$\fi}
\def \chandra  {{\it Chandra}}
\def \xmm      {{\it XMM-Newton}}

\journalinfo{The Astrophysical Journal, astro-ph/0311604}
\slugcomment{Received 2003 October 23; accepted 2003 November 27}  
\shorttitle{The UTA Feature in AGNs}
\shortauthors{H. NETZER }

\begin{document}
 
\title{The Iron Unresolved Transition Array in Active Galactic Nuclei}

\author{
Hagai Netzer,\altaffilmark{1}
}

\altaffiltext{1}{School of Physics and Astronomy, Raymond and Beverly Sackler
Faculty of Exact Sciences, Tel-Aviv University, Tel-Aviv 69978, Israel.}

\begin{abstract}                    

The unresolved transition array (UTA) of iron M-shell ions is a prominent absorption
feature in the X-ray spectrum of many
active galactic nuclei (AGNs). Modeling  
photoionized plasmas in attempt to match the 
observed silicon and oxygen lines fail to predict the level of ionization
of iron as inferred by this feature. It is suggested that the discrepancy is due to 
underestimation of the low-temperature dielectronic recombination rates for iron M-shell ions.
Modified ionization balance calculations, based on new (guessed) atomic data, support this idea. 
The results are shown and compared to
the global properties of several observed UTAs.
Implications for AGN absorbing gas
are discussed including an analysis of the ionization parameter distribution in such sources.
The need for real calculations of such atomic data is stressed.

\end{abstract}

\keywords{
galaxies: active --- 
galaxies: nuclei --- 
atomic data ---
atomic processes ---
X-rays: galaxies}

\section{Introduction}

\chandra\ and \xmm\ spectroscopy of  active galactic nuclei
(AGNs) show a rich spectrum of X-ray absorption lines superimposed on a
 power-law continuum. Common features in such spectra are H-like
and He-like lines of the  astronomically abundant elements and 
inner-shell lines of silicon, sulphur, iron and other elements
 (e.g. Kaspi et al, 2002, Behar and Netzer 2002). 
Among these, the n=2--3 inner-shell transitions of \ion{Fe}{1}-\ion{Fe}{16}
(so called iron M-shell ions) are particularly strong. They have been observed in
IRAS\,13349+2438 (Sako et al. 2001), NGC\,3783 (Kaspi et al. 2001), NGC\,5548 (Steenbrugge et al. 2003)
 and in other sources showing warm absorber (WA) features. The
lines are hard to
resolve spectroscopically  and form a broad unresolved transition array (UTA)
around 
16--17\AA. A detailed investigation of these
transitions is given in Behar, Sako \& Kahn (2001). These authors comment on the early
UTA observations, 
compute synthetic absorption spectra  and
provide abbreviated set of atomic parameters for these lines. 

The  UTA feature
has been discussed in a couple of recent papers analyzing the 900 ks
\chandra\ spectrum of NGC\,3783. Krongold et al. (2003) used the feature to
argue for a very low ionization absorber in this source while Netzer et al.
(2003, hereafter N03)  noted a disagreement
between the predicted and observed wavelength of the feature.
The equivalent width (EW) and the central wavelength of the UTA are potentially important diagnostics 
of AGN absorbers and it is important to resolve the uncertainties in the calculation of the feature.
This paper provides a new look at recently observed AGN spectra and suggests a major revision in
iron dielectronic recombination (DR) rates that can resolve the UTA problem.
\S2  describes the available data and the old calculations.
In \S3  I present new calculations of the
spectral distribution of the UTA based on ad hoc increases in the DR
rates  
and \S4 summarizes the more important new findings of this work.

\section{UTA Observations and old calculations}

The iron UTA feature has now been observed in the X-ray spectrum of at least six AGNs. The first
detection and its implications are discussed in Sako et al. (2001). These authors analyzed the
\xmm\ RGS spectrum of IRAS\,13349+2348 and noted the strong absorption feature at 16--17\AA. They
 also produced
a detailed comparison with the widths and EWs of other strong absorption lines in the spectrum of this
sources. The feature was observed in the first \chandra\ HETG spectrum of NGC\,3783 (Kaspi et al. 2001)
and confirmed, with better measurements, by the second, 900 ks spectrum of the source (Kaspi et al.
2002). It was also  observed in the RGS spectrum of this source
by  Blustin et al. (2003) and by Behar et al. (2003).  UTA features have  been detected in the RGS spectrum
of NGC\,5548 by Steenbrugge et al. (2003) and in the RGS spectrum of Mkn\,766 by Mason et al. (2003).
 More tentative identifications are mentioned by Pounds et al. (2001) who observed the
X-ray spectrum of Mkn\,509 and by Kaspi et al. (2003) who found the
signature of the feature in the RGS spectrum of the quasar MR2251-178.

To illustrate the possible range of shapes and EWs of the observed UTAs, I show in Fig. 1 the 10--20\AA\ spectrum
of four AGNs with good S/N over this wavelength range (for information on the data used see the figure caption).
 Two of the spectra show very strong UTAs.
These are the low-state spectrum of NGC\,3783 (N03) and the
spectrum of IRAS\,13349+2348
(Sako et al. 2001). The third
case is 
MCG-6-30-15. The RGS spectrum of this source contains very strong and broad features
that were interpreted  as either due to  broad relativistic emission lines 
(Branduardi-Raymont et al. 2001; Sako et al., 2003) or 
due to a dusty WA (Lee et al. 2001). Regardless of the exact model, the spectrum shows a 
dip over the wavelength range where a strong UTA is detected in the spectrum
of NGC\,3783. This is very likely due to iron M-shell absorption although a full model is required to prove
this case.
The fourth case is the \chandra\ LETG spectrum ( obtained from the \chandra\ archive) of
NGC\,5548 that also shows a shallow absorption feature over the same range.
The main conclusions of this comaprison are the great similarity in width and central wavelength
of these UTAs and the large range in the EW of the feature.

\begin{figure}
\plotone{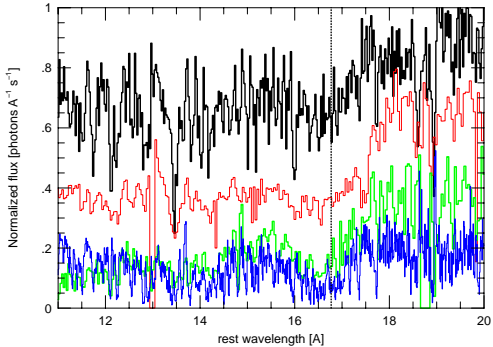}
\caption{
Normalized 10--20\AA\ \chandra\ and \xmm\ spectra of AGNs showing  UTA features.
From bottom to top:
{\bf NGC\,3783} (data from N03).
{\bf IRAS13349+2438} (RGS data from the \xmm\ archive, basically
identical to the spectrum shown in Sako et al. 2001).
{\bf MCG-6-30-15} (259 ks RGS spectrum from the \xmm\ archive).
{\bf NGC\,5548} (339 ks LETG spectrum from the \chandra\ archive).
The vertical line at 16.77\AA\ marks the location of the \ion{O}{7} b-f edge.
}
\label{fig1_label}
\end{figure}  

A detailed theoretical study of the UTA is given in Behar et al. (2001). These authors
 identify the strongest lines in the feature as due to 2p-3d transitions in iron M-shell ions 
and provide new atomic calculations (wavelengths, A values and autoionization rates) to enable
their computation. They also list abbreviated atomic data and show synthetic spectra for all
the relevant ions (\ion{Fe}{1}--\ion{Fe}{16}). Their published 
data set includes mean wavelengths,  summed
f-values, summed life-times and total widths (in \AA)
 of the main M-shell absorption features in all those ions.
The paper contains several models of photoionized plasma computed with XSTAR (Kallman and Krolik 1995)
and  pertaining to AGN absorbers. They also comment 
on various other issues such as the total absorbing column deduced from the iron M-shell lines
and the conditions for line saturation.

More recent calculations are provided by Krongold et al. (2003) and N03 who analyzed the
900 ks \chandra\ HETG spectrum of NGC\,3783. Krongold et al. modeled the spectrum by two absorbers and
 used the observed wavelength of
the UTA feature as their strongest evidence for a very low ionization parameter component.
They also used the shape of the feature to argue against
a flat ionization distribution of absorbers (see below). Their calculations are performed by
G. Ferland's
 photoionization code CLOUDY
which was supplemented by an extensive list of atomic f-values for many elements and ions.
The abbreviated Behar et al. (2001) atomic data set was used for calculation the UTA spectrum.
N03 present a more complex absorber made of three ionization components and
six kinamic components. The calculations are made with ION2003, the 2003 version of ION
 (Netzer 1996).
This code contains the full atomic data set for the iron M-shell ions (provided by E. Behar)
with many A-values and autoionization rates for each of the ions.
N03 argue for a noticable disagreement between the predicted and the observed level of
ionization of iron, as inferred from the UTA measurments (see \S3)
 and suggest that the reason 
may be the DR rates for the iron M-shell ions. The differences between the N03 and the 
K03 interpretations are much
larger than expected from the differences in the atomic data used by the two groups. This
putative discrepancy is discussed in the following sections.

\section{New calculations}
This section suggest a way to cure the discrepancy between the observations and the calculations of the 
iron lines and provided new calculations to test this idea.

\subsection{Low temperature dielectronic recombination rates for iron}

Radiative recombination (RR) and DR rates for
iron have been discussed extensively
in the literature (see Arnaud and Raymond 1992, hereafter AR92;
Mazzota et al., 1998; Savin et al. 2002; Savin et al. 2003; 
Gu 2003a, Gu 2003b). In short,
there are reasonably accurate  calculations of RR rates  for almost all iron ions
and a few laboratory measurements and
theoretical calculations for $\Delta n=1$ (so called ``high
temperature'') DR rates for iron L-shell ions.
There are also some calculations of high-temperature DR rates for iron M-shell ions.
The situation regarding
$\Delta n=0$ (``low temperature'') DR rates is more problematic. There are some estimates and several
measurements for the L-shell ions but the rates for the M-shell ions
are basically unknown and 
there are no  laboratory measurements to identify and measure the relevant
resonances. The only exception is the case of \ion{Fe}{16} where several resonances have been measured
and converted to DR rates (Linkemann et al. 1995; Muller, 1999).
 Since low-temperature DR rates in other elements are comparable
or even exceed the RR rates (Nussbaumer and Storey,
1983, 1984, 1986, 1987), this is likely to be the case for at least some iron ions ( e.g. the above
mention DR rates for \ion{Fe}{16} are orders of magnitude larger than the Arnaud and Raymond
(1992) theoretical rates).
Thus the ionization balance of iron in photoionized palasmas where M-shell ions are
the most abundant, is rather uncertain.

The older RR and DR rates for iron ions used in photoionization codes like
CLOUDY and ION  are mostly from the compilation of AR92.
Extensive tests, over a large range of density and ionization parameter show
that the new Gu (2003b) DR rates for the L-shell iron ions produce similar
fractional ionizations to the AR92  rates 
(the seemingly large deviations illustrated in Gu 2003b, Fig. 22, are
for much higher temperatures typical of collisional plasmas).  Regarding the iron
 M-shell ions, the Arnaud and Raymond (1992) compilation give  basically 
high temperature rates and  the
low-temperature recombination is controlled primarily by RR

As argued in N03, the new X-ray observations of NGC\,3783
show a discrepancy between the ionization fraction of iron and
various other elements. In particular, the level of ionization of iron
deduced from the central wavelength and the shape of the UTA feature indicates
lower ionization parameter compared with the one deduced from several
oxygen and silicon lines. The best example is the comparison of the
EWs of  \ion{Si}{9}, \ion{Si}{10} and \ion{Si}{11}
lines, near 7\AA\ with the EW and central wavelength of the iron UTA.
According to N03, 
all those features originate in the lowest
ionization component (the one with log($U_{OX})=-2.4$ where $U_{OX}$ is the ``oxygen ionization parmeter'' defined
over the 0.54--10 keV range). The above silicon lines
are well separated and easy to measure in the spectrum of NGC\,3783. Their
f-values have recently been calculated by Behar and Netzer (2002) and other
atomic data for silicon ions are reasonbaly well known. Thus, the column densities
 deduced from the measured EWs
of those ions are realiable ionization parameter measures. For the deduced 
ionization parameter,   \ion{Fe}{10}--\ion{Fe}{12} are calculated to be the dominant
iron ions. However, the UTA observation corresponds to 
\ion{Fe}{8}--\ion{Fe}{10}, i.e. a considerably more neural iron.
 A similar discrepancy is found when
comparing the oxygen level of ionization with that of
iron while the oxygen and silicon levels of ionization seem to be consistent
with each other. K03 also discussed the silicon-iron discrepancy and noted that their
best model fails to predict the observed intensities of the
\ion{Si}{10} and \ion{Si}{11} lines. 

The discrepancy in the level of ionization of the absorbing gas in NGC\,3783 cannot be cures
by changing the density, the spectral energy distribution (SED) or any other
parameters of the model. (Note that high density suppression of DR rates must have an effect
at some density which is probably well above the one deduced for NGC\,3783. Obviously such
atomic data are not yet available).  Changing the iron abundance by factors of 3--5 does
not help either. Lowering the ionization, as suggested by Krongold et al. (2003), 
improves the fit of the UTA wavelength but results in a poor fit of the silicon lines.
I therefoe suggest that the origin of the disagreement is the
underestimation of the low temperature $\Delta n=0$ DR rates for the iron M-shell
 ions. This assumption is investigated in the following section. 

\subsection{Assumed low-temperature DR rates}

The standard expression for fitting DR rate coefficients is
\begin{equation}
\alpha_{DR} = T_e^{-3/2} \sum C_i \exp {(-E_i/k T_e)} 
\end{equation}
where $C_i$ are fitting constants, $E_i$ are energies expressed  in temperature units and $T_e$ the electron
temperature. The procedure adopted here is to add a {\it single term} to the
existing formula for each of the ions  \ion{Fe}{6}--\ion{Fe}{15} with constants $C_0$
and $E_0$,  to represent a guess of the low temperature DR rate. 
The additional
terms were chosen such that $E_0/k= 3\times 10^4$~K and the new $\alpha_{DR}$ equals the
RR rate at that temperature. This results in a small increase of the total rate 
at $T_e=10^4$K and an increase by factors of 2--3 at $2\times 10^4 - 10^5$\,K.
The ratio of the new total (RR+DR) rates to the older values, at various
representing temperatures, are given in Table 1. Also listed are the chosen values of $C_0$ for relevant
iron ions.
\begin{deluxetable}{lcccc}
\tablecolumns{5}
\tablewidth{0pt}
\tablecaption{Present and AR92 RR+DR rates\tablenotemark{a}
\label{DR rates}}
\tablehead{
\colhead{Ion$/ {\rm T_e}$ } &
\colhead{ ${\rm 10^4 K}$  } &
\colhead{ ${\rm 2 \times 10^4 K}$ } &
\colhead{ ${\rm 5 \times 10^4 K}$ }&
\colhead{$C_0$}  }
\startdata
 \ion{Fe}{6}  &  1.33 & 1.87 & 1.95 & $1.8 \times 10^{-4}$  \\
 \ion{Fe}{7}  &  1.32 & 1.86 & 2.09 & $2.7 \times 10^{-4}$ \\
 \ion{Fe}{8}  &  1.31 & 1.85 & 2.09 & $3.8 \times 10^{-4}$ \\
 \ion{Fe}{9}  &  1.34 & 1.95 & 1.26 & $5.7 \times 10^{-4}$ \\
 \ion{Fe}{10} &  1.31 & 1.88 & 1.43 & $7.0 \times 10^{-4}$ \\
 \ion{Fe}{11} &  1.30 & 1.85 & 2.14 & $8.2 \times 10^{-4}$ \\
 \ion{Fe}{12} &  1.30 & 1.85 & 2.14 & $1.0 \times 10^{-3}$ \\
 \ion{Fe}{13} &  1.30 & 1.85 & 2.05 & $1.2 \times 10^{-3}$ \\
 \ion{Fe}{14} &  1.36 & 1.89 & 1.98 & $1.1 \times 10^{-3}$ \\
 \ion{Fe}{15} &  1.34 & 1.85 & 1.61 & $1.2 \times 10^{-3}$ \\
\enddata
\tablenotetext{a}{The numbers in the table give the ratio of the present to
the AR92 total recombination rates. Column 5 gives the additional constant for
the new DR rate in  equation 1.
}
\end{deluxetable}   
 
The above prescription is only one out of several possible ways to
artificially increase the low temperature DR rates. The only observational handle on such atomic
rates is related to UTA observations and different choices of the values of
 $E_0$ and $C_0$ might work just as well.
 All  prescriptions
of this type result in a large increase ($\sim 3$) of the total recombination rate at 
$T_e \simeq 10^5$K.
However, in photoionized gas, such temperatures are normally found in regions
dominated by
L-shell rather than M-shell iron ions. Thus the main effect of
the new rates is on the ionization balance in those regions where
the electron temperature is   $1-4 \times 10^4$~K.

Several models of photoionized plasma have been computed to
explore the effect of the new DR rates 
on the ionization balance of iron.
The method of calculation and all atomic data, except for the present DR rates, are similar to
the ones discussed in N03. 
 The assumed  SED  is the N03 NGC\,3783 continuum with a 0.1--50 keV 
photon  index of $\Gamma=1.8$.
The gas density is low (of order $10^5$)
cm$^{-3}$ such that collisional excitation of high levels are negligible and the gas composition
is the one given in N03 (Table 2).
  The chosen hydrogen column density is $10^{21.5}$ \cmii\ and the turbulent velocity 250 \kms.
This combination was chosen to represent those cases that show clear iron absorption lines and
\ion{O}{7} absorption edge. Saturation of the strongest iron lines can become important at
such column densities. This was taken into account in the line excitation process but the
calculated line profiles were assumed to be Gaussian.

Fig. 2 shows the
optical depth of the leading  lines (those with the largest
absorption cross section) of several iron M-shell ions as a function of $U_{OX}$ for the present
and the AR92 DR rates.
Also shown is the optical depth of  \ion{O}{7} bound-free  
 edge at 16.77\AA.
The use of the present rates has a significant influence on  iron 
whose mean level of ionization is lowered by one or two ions
for a given  ionization parameter.
The mean increase in log($U_{OX}$) required to obtain the
level of ionization of iron obtained with the AR92 rates is about 0.2.
\begin{figure}
\plotone{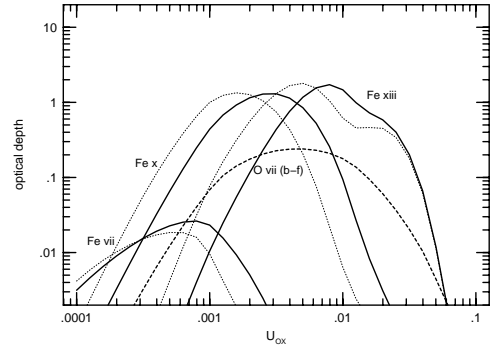}
\caption{
Optical depths of the leading (largest cross section) \ion{Fe}{7}, \ion{Fe}{10} and
\ion{Fe}{13}  lines and
the \ion{O}{7}  bound-free  absorption edge as a function of the ionization
parameter. Solid line: present DR rates. Dotted line: AR92 DR rates.
Other model parameters are given in \S3.2.
}
\label{fig2_label}
\end{figure}  

The present DR rates were also used to re-compute a full model for NGC\,3783 using the
six absorption components and the three emission components described in N03.
Fig. 3 shows a comparison of the old and the new
models with the low-state spectrum of the source (N03 Fig. 7). The diagram shows that
the increased DR rates cause a significant shift in
iron ionization and a much improved fit to the data.  The level of
ionization of iron, oxygen and silicon are now all consistent and the 
present rates  seem to cure the ionization discrepancy (the silicon line fit over
 the 4--7\AA\ band has not changed and 
is basically identical to the one shown in N03).
Applications to the spectrum of other sources are more difficult to illustrate
since we do not have satisfactory models for these cases.
\begin{figure*}
\hglue-0.4cm{\includegraphics[angle=0,width=10cm]{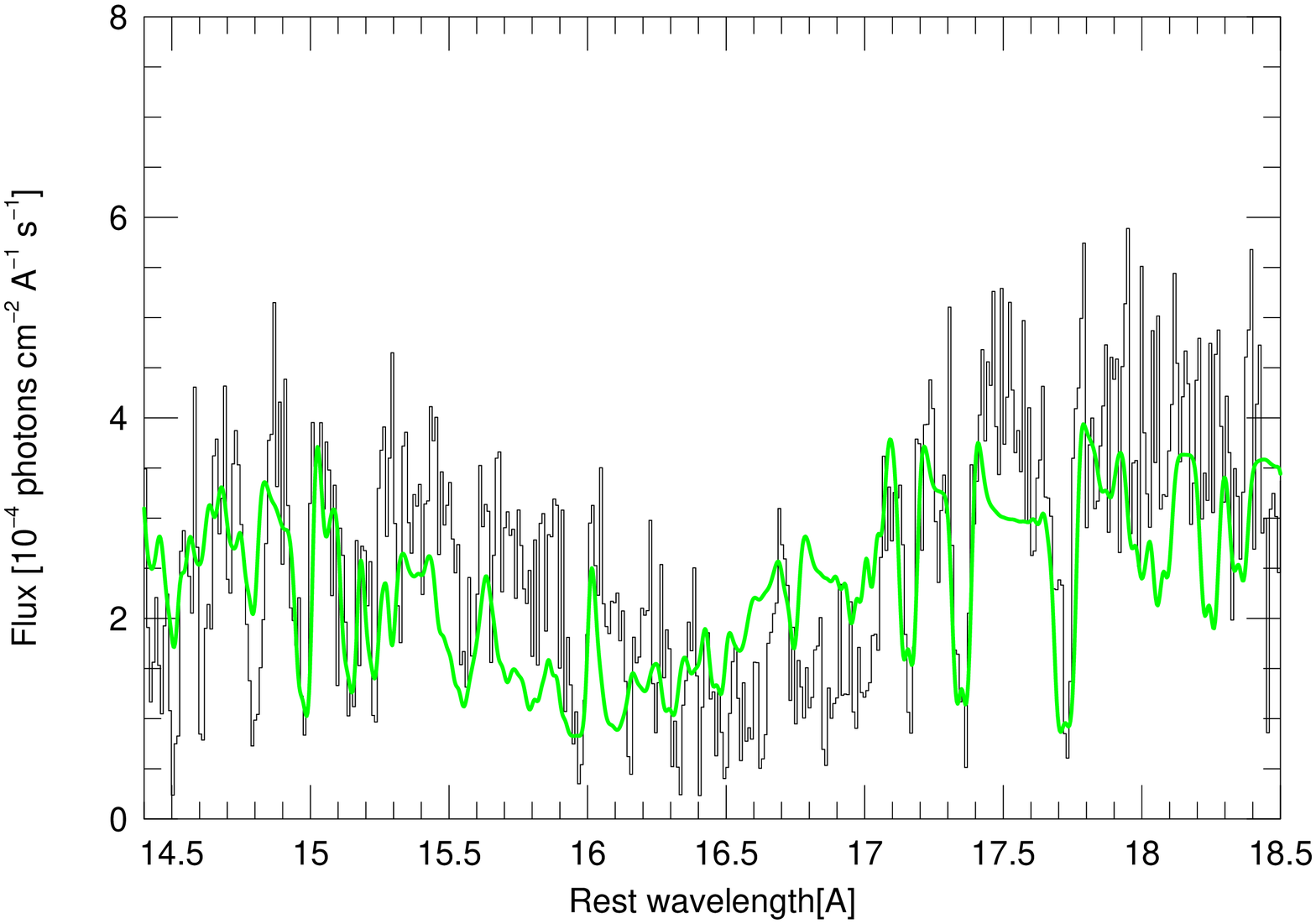}}
\hglue-0.8cm{\includegraphics[angle=0,width=10cm]{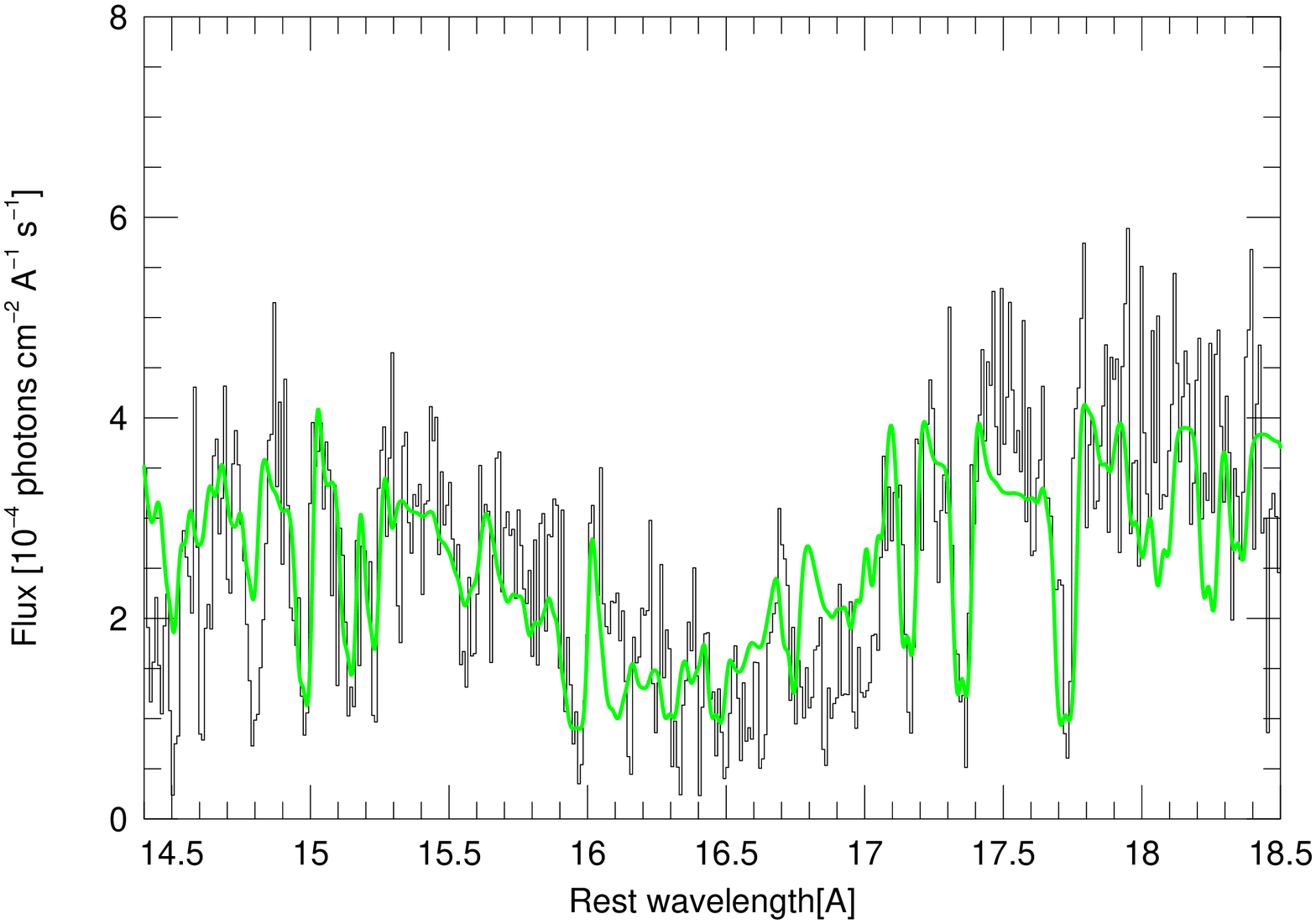}} 
\caption{Left: Old fit to the low-state spectrum of NGC\,3783 (from N03 Fig.
7).  Right: Same model components but using the new DR rates.
}
\label{fig3_label}
\end{figure*}

\subsection{Discrete versus continuous distribution of ionization}

The Behar et al. (2001) work  demonstrates the usefulness of the UTA
observations in  determining the level of ionization of the X-ray absorbing gas. In
particular, the central wavelength of the feature is an indicator of the
dominant iron ions. Thus,
accurate UTA measurements can be used to test
the idea, raised by Kinkhabwala et al. (2002) who studied the spectrum of the type II Seyfert NGC\,1068,
that the highly ionized emitting gas in AGNs
indicates a flat distribution of ionization rather than a few discrete components.
If type-I AGN absorbers are similar in their properties to type-II AGN emitters,
 this would  mean that in such absorbers, many ions of the same
element have comparable column densities.
 Such an ionization structure can be formed 
either due to a strong dependence of the ionization parameter on the distance from the center  or due to
multi-component multi-density medium at a given location (which is more likely to be the
case for NGC\,1068).
A similar idea, by Krolik and Kriss (2001), involves a medium with a smooth distribution of
temperatures and ionization parameters. 
The predicted spectroscopic differences  are clear. A single ionization component will result in a relatively
narrow UTA while the suggested flat distribution will produce a much broader feature.

The idea of a flat ionization distribution was further investigated by
Krongold et al. (2003).
These authors provided several calculations similar to those shown by 
 Behar et al. (2001) to support the idea of a broader UTA for larger ionization parameter gas.
They also argued that the sharply defined and relatively narrow UTA observed in the spectrum of NGC\,3783
 is inconsistent with the Krolik and Kriss (2001) suggestion since any  mixture of
ionizations will results in a broader and shallower feature.

To further test this suggestion, I calculated several multi-components photoionization models  
 and compared them with 
``single shell'' models. The calculations employ the new DR rates for the iron M-shell ions.
One example, shown in Fig. 4 compares 
the combined spectrum of four absorbing  shells spanning  a large
range of ionizations (a factor of 10$^{1.5}$ in $U_{OX}$) with 
the spectrum of a single shell whose column density equals the sum
of the four columns.
 The column density of each shell was
10$^{21}$ \cmii\ and the  four ionization parameters were chosen such that
in one shell \ion{O}{5} is the most abundant oxygen ion, in the
second \ion{O}{6} is most abundant, \ion{O}{7} dominates  the third shell ionization and
\ion{O}{8} the fourth. The total column densities of those ions, in all
four shells, are: 10$^{17.54}$ \cmii\ for \ion{O}{5},
10$^{17.56}$ \cmii\ for \ion{O}{6},
10$^{17.80}$ \cmii\ for \ion{O}{7},
and 10$^{17.52}$ \cmii\ for \ion{O}{8} (note that it is difficult to obtain a 
more uniform distribution because several oxygen ions are present in each shell).
The single-shell model column density is 
10$^{21.6}$ \cmii\ (identical to the column density of the low ionization component of 
Krongold et al. 2003), and log$(U_{OX})=-2.5$, roughly the mean of the four ionization parameters
and similar to the lowest ionization parameter component in NGC\,3783.
\begin{figure}
\plotone{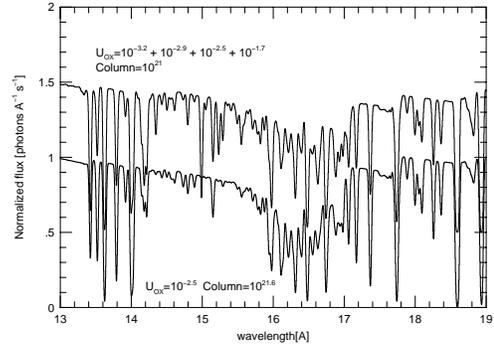}
\figcaption {A comparison of a four component model with ionization parameters
and column densities as marked (top curve) with a single component model with
a column density of the total four columns and a ``mean'' ionization parameter
(bottom curve). The wavelength covered by the UTA under such conditions is roughly
15.2--17.4\AA. The column density and ionization parameter of the four components are
marked at the top and those for the single shell at the bottom of the diagram.
Note the great similarity in total width and central feature of the calculated UTAs.
\label{fig4_label} }
\end{figure} 

As shown in Fig. 4, the composite four shell spectrum and the
single-shell spectrum produce very similar UTAs.
The reason is that the thick single shell contains a mixture of oxygen ions which is
not too different from the four-shell mixture. This is also
true for the M-shell iron ions whose total column densities in the four-component are
comparable to the column densities in the single shell.
In fact, the only significant difference between the two is the
presence of a strong \ion{Fe}{17} absorption line, near 15\AA, in the four-component model. This is caused
by the larger column density of high ionization iron in the four component spectrum.

The above result is very general and other combinations of somewhat smaller or larger
column density shells lead to similar conclusions.
In general, thicker shells contain broader distribution of ionizations and produce 
broader UTAs. As noted by Behar et al. (2001), and Krongold et al. (2003), higher mean ionization also
result in broader UTAs yet all those single-shell cases can be broken into several sub components
with a similar combined spectrum.
Thus, UTA observations by themselves cannot be used to rule out the
possibility of a multi-component, flat ionization parameter distribution of absorbers
spanning 1--2 orders of magnitude in ionization parameter. This does not seem to be a consequence
of the new DR rates since adding those rates result in roughly a constant shift of all ionization
fractions as a function of $U_{OX}$ (Fig. 2).

\section{Conclusions}

Detailed analysis and modeling of  iron UTA features in AGN spectra suggest
that the iron level of ionization is inconsistent with the ionization of
other elements. It is suggested that the reasons are the inaccurate DR rates used for 
the iron M-shell ions and new "assumed" rates are used to support this claim.
Theoretical spectra using the new atomic data give much better fit to the
observed spectrum of NGC\,3783.
It is also shown that UTA observations by themselves cannot be used
to distinguish between multi-component and single component low ionization AGN absorbers.
New measurements and calculations of DR rates for M-shell lines are badly needed to produce more
realistic models to the X-ray spectrum of AGN.

\acknowledgements
I am grateful to Daniel Savin and Ehud Behar for interesting and useful discussions.
I also thank the referee, D. Liedahl, for his many useful suggestions.
Shai Kaspi  provided his great expertise in reducing
the \chandra\ and \xmm\ archival spectra.
 This work is supported by the Israel Science
Foundation grants 545/00 and 232/03. I thank the astrophysics group at Columbia
university for their hospitality during a summer visit in 2003.

\end{document}